# "Using routinely collected patient data to support clinical trials research in accountable care organizations"

**Authors**
[1] Andrew J McMurry, PhD Corresponding author (andy.mcmurry@medal.com)

[1] Richen Zhang, MS

[2] Alex Foxman, MD, FACP

[2] Lawrence Reiter, PhD

[2] Ronny Schnel, MA

[1] DeLeys Brandman, MD, MPH

**Affiliations**
[1] Medal, Inc ( HITRUST certified Health-IT company)

[2] NACORS, LLC (National Accountable Care Organization Research Services)

## ABSTRACT

**Background**: More than half (57%) of pharma clinical research spend is in support of clinical trials. One reason is that Electronic Health Record (EHR) systems and HIPAA privacy rules often limit how broadly patient information can be shared, resulting in laborious human efforts to manually collect, de-identify, and summarize patient information for use in clinical studies.

**Purpose:** Conduct feasibility study for a Rheumatoid Arthritis (RA) clinical trial in an Accountable Care Organization. Measure prevalence of RA and related conditions matching study criteria. Evaluate automation of patient de-identification and summarization to support patient cohort development for clinical studies.

**Methods:** Collect original clinical documentation directly from the provider EHR system and extract clinical concepts necessary for matching study criteria. Automatically de-identify Protected Health Information (PHI) protect patient privacy and promote sharing. Leverage existing physician expert knowledge sources to enable analysis of patient populations.

**Results:** Prevalence of RA was four percent (4%) in the study population (mean age 53 years, 52% female, 48% male). Clinical documentation for 3500 patient were extracted from three (3) EHR systems. Grouped diagnosis codes revealed high prevalence of diabetes and diseases of the circulatory system, as expected. De-identification accurately removed 99% of PHI identifiers with 99% sensitivity and 99% specificity.

**Conclusions:** Results suggest the approach can improve automation and accelerate planning and construction of new clinical studies in the ACO setting. De-identification accuracy was better than previously approved requirements defined by four (4) hospital Institutional Review Boards.

## I. INTRODUCTION

Clinical trials require both sufficient numbers of human samples and samples from different patient populations -- a single large medical center is not sufficient no matter the size [1]. Differences in patient demographics, disease prevalence, healthcare settings, coding practices, and health information systems all limit how broadly study results will demonstrate success for patients outside of the study population [1]. To defer difficult reproducibility issues to phase III clinical trials would add significant time and cost.

Clinical research spending is increasing year-on-year with fewer new drugs being brought to market. More than half (57%) of pharma clinical research spend is in support of clinical trials [2], and many clinical trials fail despite promising early findings. Nonetheless, the industry of drug discovery continues to grow continues to grow and is supported by Health-IT applications [3] for traditional clinical studies [4] and modern precision medicine [5].

This paper reports methods and results of extracting patient data to support clinical research [6], cohort selection, and clinical trials in diverse ACO locations. The methodology reported here previously aided the query and analysis of patients with Juvenile Idiopathic Arthritis (JIA) across 60 health institutions [7], suggesting that the approach may be applicable to adults with arthritis as well.  The methodology is then validated in three real world settings to identify a target patient population diagnosed with rheumatoid arthritis (RA) and related diagnosis.

## II. METHODS

### Patient Data Extraction For Cohort Analysis
Patient data from Electronic Health Record (EHR) systems are extracted and translated into U.S. standard coded value sets [8] for demographics [9, 10], diagnosis [11], medications [12], and procedures [13]. To support population level analysis [1] [14] , coded concepts were grouped using expert hierarchies curated by physician experts, shown in Table 1.

| PATIENT DATA | CODING STANDARD | EXPERT GROUPING |
|---|---|---|
| Diagnoses | ICD (ICD-9-CM, ICD-10-CM) SNOMED Clinical Terms | CCS2 Clinical Classifications UMLS Metathesaurus |
| Medications | RxNorm NDC | NDF-RT Drug Ingredients UMLS Metathesaurus |
| Lab Tests | LOINC | LOINC Test Panels |
| Demographics | HL7 Core | CDC Race and Ethnicity |

**Table 1.** *Clinical data standardization with clinical codes grouped by clinical expert knowledge.* Diagnoses are extracted using SNOMED Clinical Terms or ICD codes. For historical diagnosis, ICD-9-CM terms are used and for recent diagnosis, ICD-10-CM terms are used. CCS Clinical Classifications and UMLS are used to group patients having diagnosis codes with synonymous or more specific meaning. Medications are extracted using RxNorm and NDC codes and mapped using NDF-RT into higher level groups such as "Antirheumatic Agents" (NDFRT:MS100). Lab Tests are indexed using LOINC standards and grouped according to Lab Test panels such as "Basic Metabolic Panel" or "Complete Blood Count." Demographics are defined by HL7 and patients are grouped using CDC Race and Ethnicity code sets. Other types of clinical data such as surgical procedures, allergies, and immunizations are standardized and grouped using the same approach.

Patient data are extracted from EHR systems to support patient cohort selection and population analysis (**Figure 1**). Medal used existing EHR interfaces without customization or need for custom and cost intensive integration -- especially in healthcare settings where the EHR system does not readily provide downloads.

Medal provides a method to replace fax-based record sharing with a robust Natural Language Processing (NLP) system to extract machine readable coded concepts from existing EHR data. Medal also supports existing HL7 standards for clinical documentation architecture, especially Continuity of Care Documents (CCD) [15], Application Programming Interfaces (API), and HL7 Fast Healthcare Interoperability Resources (HL7 FHIR) [16].

Patient data are HIPAA de-identified to enable sharing the original clinical documentation between clinical research organizations. Patients are grouped using expert knowledge systems to enable query and analysis of higher level disease categories such as "*Diseases of the musculoskeletal system and connective tissue*" [17] and medications "*Antirheumatic Agents*" [18].

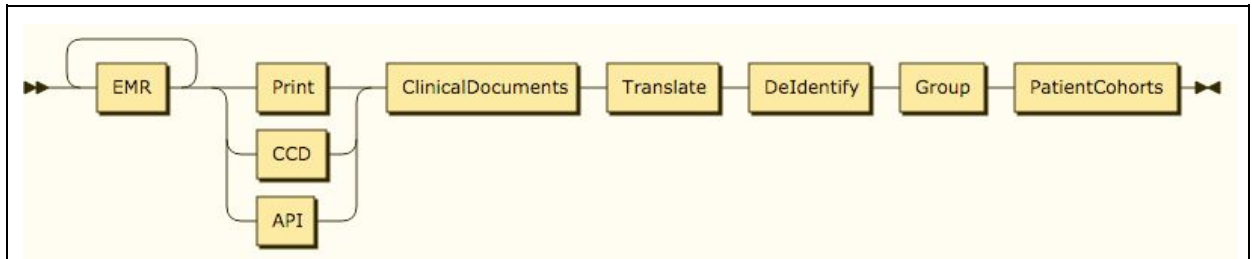

**Figure 1: Patient data extraction for patient cohort analysis.** Clinical documents from existing EHRs are extracted using Print, CCD, and API methods. Clinical documents are then translated into standards, de-identified, and grouped using expert knowledge sources. Patient cohorts can then be easily selected and analyzed to support a wide range of clinical research use cases.

## III. RESULTS

Patient data were extracted and de-identified from three EHR systems (Practice Fusion, GE Centricity, and drchrono) for NACORS care providers. In total, records were extracted for 3,500 NACORS patients (**Figure 2**). The average patient age was 53 years old and the data reflected a nearly even division of female (52%) and male (48%) patients (**Figure 3**). The most common diagnoses were "*Vitamin D Deficiency* "*Pure Hypercholesterolemia,*" and "*Essential Hypertension.*" Grouping patient observations using expert knowledge sources reflected known disease burdens in adult populations, namely diabetes and heart disease (endocrine disorders and diseases of the circulatory system, respectively, **Table 2**). Rheumatoid Arthritis (RA) was chosen as the case study for patient cohort selection to support clinical trials. RA standard case definition was used to select patients with RA and patients with diseases similar to RA, reflecting the needs to support clinical trials with covariates, outliers, co-morbidity [14] , and patients who may have RA but not yet had a diagnosis. Prevalence of RA was ~4% in the study population. The results suggest the approach is useful and quickly applicable to many use cases within clinical research [1].

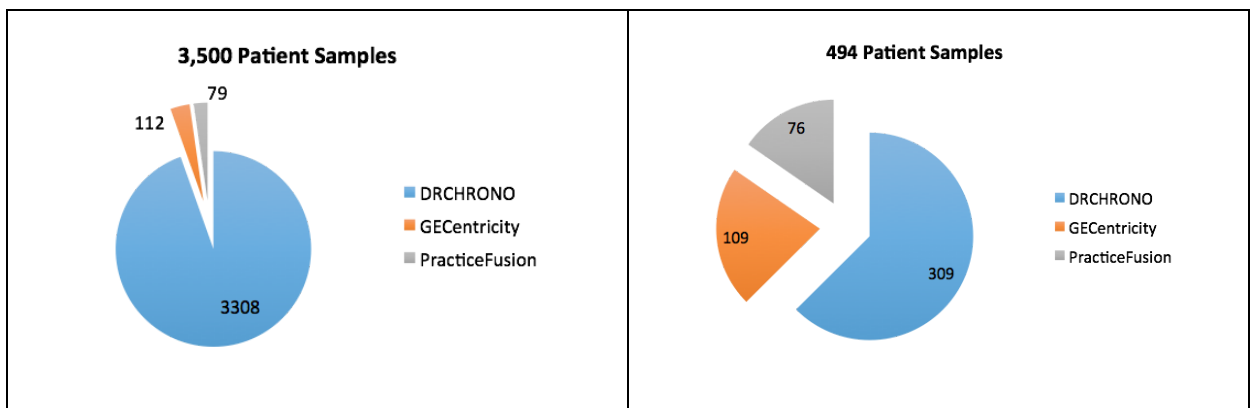

**Figure 2: Patient samples extracted from three EHR systems.** Left: 3,500 patient samples were extracted from Practice Fusion, GE Centricity, and drchrono. Right: 494 patient samples included at least one visit diagnosis.

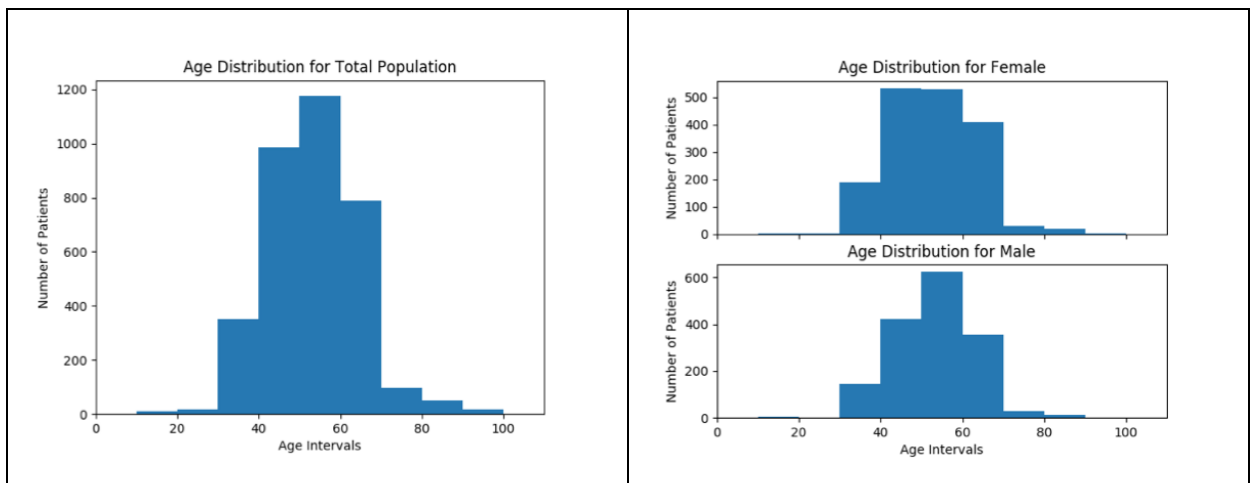

**Figure 3: Demographic breakdown for age and sex.** Left: Average age for all patients was 53 years old. Right: Age distribution by sex. There are roughly equal numbers of females and males in this dataset (52% vs 48%). The age distribution in the sample population for is slightly older for men.

## Disease Prevalence

Rheumatoid Arthritis (RA) was chosen as the case study for patient cohort selection to support clinical trials. RA standard case definition was used to select patients with RA and patients with diseases similar to RA, reflecting the needs to support clinical trials with covariates, outliers, co-morbidity, and patients who may have RA but not yet had a diagnosis.

Rheumatoid Arthritis is defined by the US National Library of Medicine as "*a form of arthritis that causes pain, swelling, stiffness, and loss of function in your joints. It can affect any joint but is* *common in the wrist and fingers* [19]". To quantify the number of patients with RA and related diseases, diagnosis codes were extracted and analyzed using SNOMED Clinical Terms and the International Classification of Diseases (ICD). Cases of RA were defined as any physician documentation related to the SNOMED Clinical Term "Rheumatoid Arthritis", including related concepts more specific than SNOMED codes 156471009 or 69896004. For ICD-10-CM diagnosis, "Rheumatoid Arthritis with rheumatoid factor"[20] provided 281 diagnosis codes. To select groups of patients with RA and related diseases, patients were combined into a larger group in the category "Diseases of the musculoskeletal system and connective tissue", defined by the expert knowledge source "Clinical Classification System".

The sample adult population had high rates (61%) of "*Endocrine; nutritional; and metabolic diseases and immunity disorders*", mostly due to diabetes and associated co-morbidities. Similarly, about half (47%) of patients had at least one diagnosis in the category "*Diseases of the circulatory system*". Relevant to the study of RA, about one third (36%) of patients had a diagnosis in the category "*Diseases of the musculoskeletal system and connective tissue.*"

| PERCENTAGE | DIAGNOSIS CATEGORY |
|---|---|
| 61% | Endocrine; nutritional; and metabolic diseases and immunity disorders |
| 47% | Diseases of the circulatory system |
| 39% | Diseases of the respiratory system |
| 36% | Diseases of the musculoskeletal system and connective tissue |
| 33% | Symptoms; signs; and ill-defined conditions and factors influencing health status |
| 31% | Diseases of the digestive system |
| 31% | Nutritional deficiencies |
| 29% | Mental Illness |
| 28% | Diseases of the nervous system and sense organs |
| 28% | Diseases of the genitourinary system |
| 23% | Disorders of lipid metabolism |
| 19% | Diseases of the blood and blood-forming organs |
| 18% | Other nutritional; endocrine; and metabolic disorders |
| 17% | Other non-traumatic joint disorders |
| 17% | Medical examination/evaluation |
| 16% | Essential hypertension |

**Table 2: Prevalence of top 15 diagnosis groups for NACORS sample patient population.** Prevalence is the percentage of patients who have at least one recorded diagnosis in the diagnosis group defined by expert knowledge sources (Unified Medical Language System and Clinical Classification System).

**Patient Privacy**

Medal signed Business Associates Agreements (BAA) authorizing the collection, de-identification, and analysis of patient data from NACORS physician locations. The BAA authorized collection of patient data, including Protected Health Information (PHI). Medal de-identification (DEID) methods achieve industry best practice compliance and security. HIPAA defined Limited Data Sets (LDS) allows for sharing of patient data that can be linked back [21] to original sources, making it possible to identify and consent patients for clinical trials through the primary care physician. Human expert review was conducted on a sample of 1,224 Protected Health Identifiers within the extracted clinical documentation, resulting in DEID 98.7% sensitivity and 99.1% specificity. De-identification accuracy was better than previously approved requirements defined by four (4) hospital Institutional Review Boards [22].

**Figure 4. HIPAA De-identification example.** Protected Health Identifiers were extracted and redacted from clinical documents. PHI and DEID records are stored separately and linked using the policies defined by HIPAA Limited Data Set (LDS). Human review confirmed that diagnosis and other clinical concepts were preserved and not redacted from de-identified records. Model performance from a sample of 1,224 identifier tokens was 98.7% sensitivity and 99.1 specificity.

## DISCUSSION

Patient data from routine clinical encounters represent a vast and underutilized resource to support clinical research including trials in patients [1, 2]. Over the last decade, adoption of Electronic Health Record (EHR) systems has approached nearly 100% in the United States [24]. Despite this remarkable achievement, the ability to share and analyze patient data across healthcare organizations is often limited.

Variation in healthcare settings and EHR systems often limits how easily patient data can be shared and commonly analyzed [1]. In general, physicians on a single EHR system at a large Academic Medical Center (AMC) can share patient data more easily with one another than providers on different EHR systems in an Independent Physicians Association (IPA) or Accountable Care Organization (ACO). As a result, providers in ACO locations often continue to rely on the fax machine, the lowest common denominator for patient data sharing. Alternatives to the fax machine range from structured EHR integrations to manual human labor, carrying heavy financial and administrative burdens.

In order to support clinical trials of RA, multiple physicians and patient populations are needed to share patient samples in sufficient numbers for clinical investigation. RA affects less than 1 percent ( <1% ) of patients in the United States  and at varying rates in different patient groups [25]. Prior studies report RA disease prevalence varying by sex [26], age, race, and location, making it all the more pressing to obtain patient samples across different healthcare settings [25]. The approach reported here was useful in the creation of a patient registry for Juvenile Idiopathic Arthritis with 60 health institutions [7].

Prior work in clinical trials support has primarily focused on large medical centers using complex EHR systems that are generally not available to small provider practices in ACO settings. Nevertheless, the need to engage patients for clinical trials through a Primary Care Physician (PCP) remains paramount [2]. For clinical trial site selection phase, it may be sufficient to select patient cohorts from billing and administrative records. Once the patient population has been identified, it is then necessary to examine the complete patient history, which is typically only available in the original EHR clinical documentation, including the physician notes.

In this study, the authors extracted, de-identified, and summarized records from three (3) EMR systems for a study population of 3,500 patients. Queries were performed to analyze patient demographics and inclusion/exclusion criteria for clinical studies of RA (rheumatoid arthritis). Patients matching the strict definition for RA were discovered as well as patients with similar disease characteristics such as comorbidities relevant to the study of RA. The results suggest that this approach is reproducible to many disease areas and can be performed quickly and inexpensively for clinical studies.